\documentclass[utf8,sort&compress]{frontiersFPHY} 

\setcitestyle{square} 
\usepackage{url,hyperref,lineno,microtype}
\usepackage{cleveref}
\usepackage[onehalfspacing]{setspace}

\usepackage{amsmath,amssymb,amsthm,mathtools,bm,siunitx,overpic}

\newcommand{\botry}{\textit{Botryllus}}
\newcommand{\botryschloss}{\textit{Botryllus schlosseri}}
\newcommand{\bschloss}{\textit{B. schlosseri}}

\newcommand{\dt}{\delta t}
\newcommand{\thetasplit}{\theta^{\textrm{split}}}
\newcommand{\thetaint}{\theta^{\textrm{period}}}
\newcommand{\field}{\phi}
\newcommand{\basisfun}{f}
\newcommand{\weight}{w}
\newcommand{\abs}[1]{\left\lvert #1 \right\rvert}

\newcommand{\avg}[1]{\left\langle#1\right\rangle}

\def\keyFont{\fontsize{8}{11}\helveticabold }
\def\firstAuthorLast{Walker {et~al.}} 
\def\Authors{Benjamin J. Walker\,$^{1,2,*}$ and Adriana T. Dawes\,$^{3,4}$}
\def\Address{$^{1}$Department of Mathematics, University College London, London, UK\\
$^{2}$Wolfson Centre for Mathematical Biology, Mathematical Institute, University of Oxford, Oxford, UK\\
$^{3}$Department of Mathematics, The Ohio State University, Columbus, Ohio, USA\\
$^{4}$Department of Molecular Genetics, The Ohio State University, Columbus, Ohio, USA}
\def\corrAuthor{Benjamin J. Walker} 

\def\corrEmail{benjamin.walker@ucl.ac.uk}

\begin{document}
\onecolumn
\firstpage{1}

\title[Modelling mechanically dominated vasculature development]{Modelling mechanically dominated vasculature development}

\author[\firstAuthorLast ]{\Authors} 
\address{} 
\correspondance{} 

\extraAuth{}


\maketitle

\begin{abstract}
\section{}
Vascular networks play a key role in the development, function, and survival of many organisms, facilitating transport of nutrients and other critical factors within and between systems. The development of these vessel networks has been thoroughly explored in a variety of \textit{in vivo, in vitro} and \textit{in silico} contexts. However, the role of interactions between the growing vasculature and its environment remains largely unresolved, particularly concerning mechanical effects. Motivated by this gap in understanding, we develop a computational framework that is tailored to exploring the role of the mechanical environment on the formation of vascular networks. Here, we describe, document, implement, and explore an agent-based modelling framework, resolving the growth of individual vessels and seeking to capture phenomenology and intuitive qualitative mechanisms. In our explorations, we demonstrate that such a model can successfully reproduce familiar network structures, whilst highlighting the roles that mechanical influences could play in vascular development. For instance, we illustrate how an external substrate could act as an effective shared memory for the periodic regrowth of vasculature. We also observe the emergence of a nuanced collective behaviour and clustered vessel growth, which results from mechanical characteristics of the external environment.
\tiny
 \keyFont{ \section{Keywords:} Vasculature; Agent-based modelling; Mechanical feedback; Network remodelling; Botryllus schlosseri} 
\end{abstract}

\section{Introduction}

The blood vasculature system consists of a network of interconnected tissues that are required to transport nutrients to all parts of an organism, in addition to moving waste products to other organs for absorption or excretion. 
The vasculature system develops through directed differentiation of precursor cells during embryogenesis \citep{Tomanek1996}, a process also referred to as angiogenesis, and is a key developmental process in organisms ranging from invertebrates such as \textit{Botryllus schlosseri} \citep{Rodriguez2019} to humans \citep{Tomanek1996}.
In addition to the general vasculature system distributed throughout an organism, specialised vasculature is required for the proper functioning of many organs including the retina \citep{Selvam2018}, the pancreas \citep{Henry2019}, and the lining of the uterus \citep{Salamonsen2021} in humans.
The vasculature system is highly dynamic, undergoing remodelling as a result of ageing \citep{Donato2018}, wound healing \citep{Jarvinen2007}, as well as monthly uterine vascular extension and retraction associated with the menstrual cycle \citep{Salamonsen2021}.
As with many developmental processes, misregulation of angiogenesis and vasculature remodelling is also associated with disease processes, from impaired wound healing in diabetic patients \citep{Okonkwo2020} to increased nutrient supply to cancerous tumours \citep{Lugano2020,Forster2017}.
In both healthy and pathological contexts, angiogenesis and vasculature remodelling occur in a complex environment, requiring the developing vasculature to interact with and potentially modify its external environment.
Despite the importance of angiogenesis and vasculature remodelling in development and disease processes, very little is known about the nature of the interaction of the developing vasculature with its surrounding environment, including how feedback between the vasculature and its mechanical environment dictates network connectivity and structure during regrowth and remodelling.


A number of theoretical models utilising an array of analytic and simulation frameworks have been proposed to investigate vasculature development.
Continuum models of vasculature development typically consist of coupled systems of differential equations, tracking the behaviour of cells at the population level.
Such models have provided insights into vasculature development under wild-type conditions in the retina \cite{Aubert2011,Maggelakis1996} and during wound healing \cite{Flegg2012}, revealing the importance of nutrient availability (including growth factors and oxygen) in vasculature development.
By design, continuum models focus on large scale behaviours and structures involving many thousands, and often millions, of cells, and emergent model behaviours are amenable to analysis using many of the tools from differential equations and dynamical systems theory.
However, continuum models rely on averaging the behaviour of collections of cells, making it difficult to understand the role of individual cells or features, and they are not effective at capturing rare or small scale events.

A popular alternative to continuum models are individual or agent based models (ABMs), which specify rules for interacting agents, usually cells, and their emergent behaviours in the formation, maintenance, and remodelling of the vasculature. 
ABM models have been used to investigate general vasculature development in physiological conditions \cite{Bentley2008} and \textit{in vitro} \cite{Artel2011} and have complimented continuum approaches for studying development in the retina \cite{McDougall2012}.
ABMs are particularly effective at elucidating the interactions that give rise to small-scale or initial vascularisation when there may be few cells. 
For instance, fine network structure, rare events, and initiation of vasculature development are best captured by ABMs. However, the rules required for this modelling approach are often complex, with intricate interconnected systems, making it challenging to identify causal factors of emergent behaviours. 
Hence, with the fundamental drivers of network formation not yet fully elucidated, a key aim of this study will be to develop a minimal and justified ABM for vascular remodelling, incorporating effects at the phenomenological level and seeking simplicity and efficiency over quantitative faithfulness and surplus complexity.

An example application of both agent based and continuum modelling approaches is the study of vasculature development and angiogenesis in the context of cancer.
Since cancerous tumours require nutrients to support their growth, tumours have co-opted the normal physiological process to promote new vasculature formation and divert oxygen and other nutrients to the growing tumour.
Given the large number of cells in tumours, continuum approaches have been successfully applied in this context, yielding improved understanding of the role of tissue structure in glioblastoma invasion \cite{Conte2021} and tumour responses to various therapeutic interventions \cite{Hormuth2021}.
ABMs have also yielded insights into vasculature remodelling in response to growing tumours \cite{Alarcon2003}. Research in this context has also motivated the development and use of hybrid schemes, which seek to exploit the best features of both continuum and agent-based modelling frameworks. Recently, these models have yielded insights into the responses of tumours to interventions, for instance, informing clinical practices such as the dosage and timing of radiation and chemotherapy \cite{Fernandez-Romero2022,Perfahl2017,Chamseddine2020,Powathil2015,Scott2016}, which highlight the significant and developing utility of hybrid modelling approaches in the study of vasculature.

\begin{figure}
    \centering
    \includegraphics[width=0.7\textwidth]{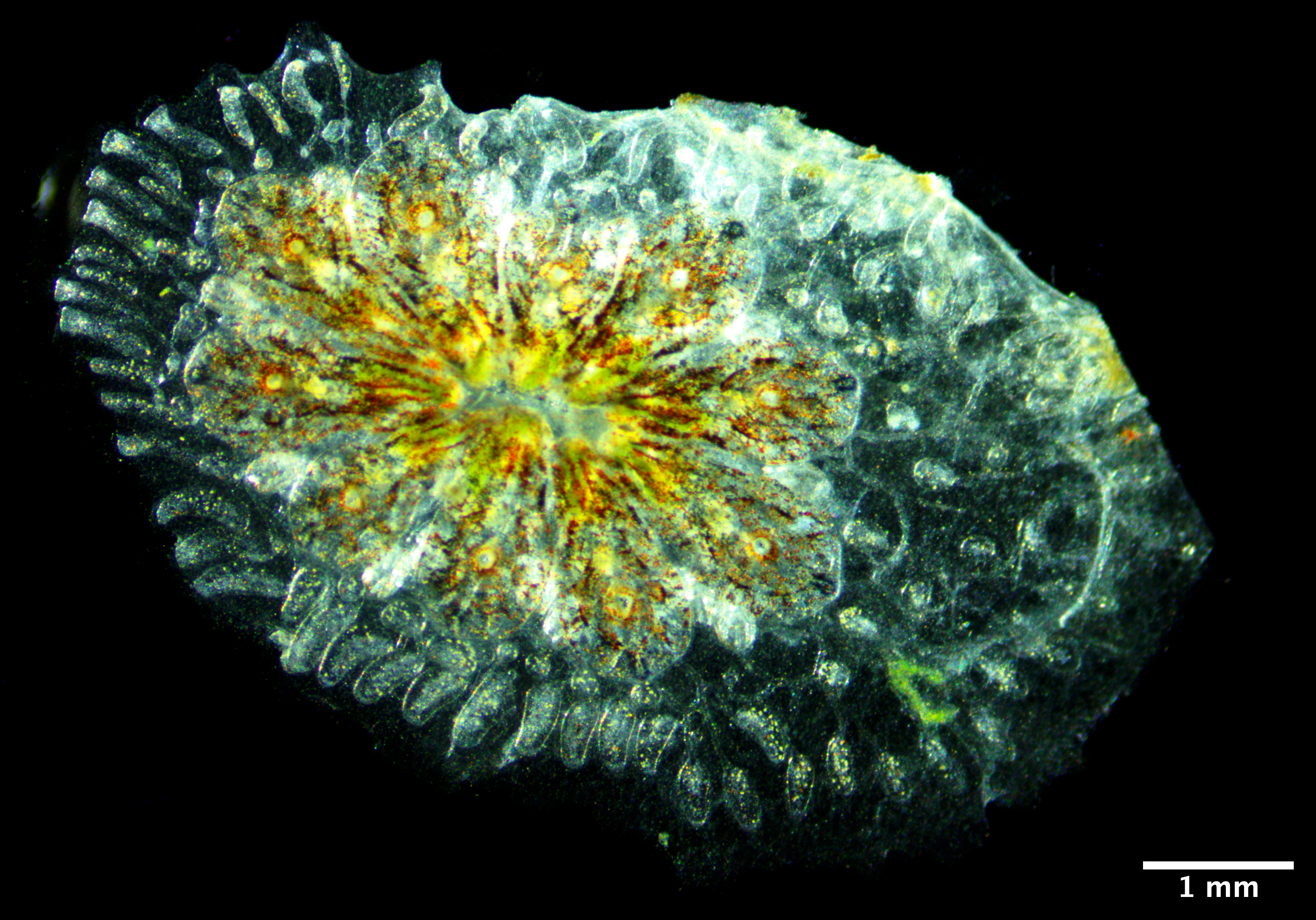}
    \caption{A typical colony of \botryschloss{}. The flower-shaped system at the centre of the organism is made up of a collection of individual zooids, with their orientation resulting in an approximate radial symmetry. The collagen-rich tunic can be seen as a grey shadow surrounding the system, and the vasculature connecting all individuals in the system can be seen as a network of vessels throughout the tunic. Image credit: Younghoon Kwon, personal communication.}
    \label{fig: botryllus}
\end{figure}

To date, the majority of models of vascular development in both wild-type and cancer contexts are driven by a chemotactic response, with vessels moving in response to gradients of growth factors \cite{Scianna2013}. There are, however, settings in which chemical effects are thought to play a secondary role in guiding network development. For example, the tunicate and model organism \botryschloss{}, shown in \cref{fig: botryllus} (Image credit: Younghoon Kwon, personal communication) and summarised in the extensive review of \citet{Rodriguez2019}, has a network of extracorporeal vessels that is exposed to the surrounding environment, which limits the potential efficacy of external diffusive species as controllers of vascular development, though in-vessel factors may still promote growth \citep{Tiozzo2008}. With the effects of chemical response less significant in this context, this motivates the consideration of other factors in vascular network formation. As noted in \citeauthor{Rodriguez2019}'s review, \bschloss{} also serves to motivate consideration of a particular class of effects, specifically the mechanical influence of a substrate, with the vessel system of \bschloss{} developing within a thin tunic \citep{Tiozzo2008}. Whilst previous studies have considered mechanical influences in other systems \citep{Murray1983,Manoussaki1996,Tosin2006,Perfahl2017,Dyson2016}, the precise nature of any interactions between this tunic bed and the growing vessels is unknown. However, it is reasonable to suppose that the extracellular medium both modifies and is modified by the movement of the vasculature, at both local and non-local scales. Motivated by this potential prominence of distributed mechanical effects, the exploration of the behaviours that emerge from such a coupling forms the primary aim of this study. Specifically, we will seek to highlight how mechanically driven growth can give rise to dynamics that are qualitatively distinct from those typically associated with chemically regulated development. Throughout, we will focus on phenomenological exploration, aiming to develop intuition via a flexible qualitative model rather than establishing quantitative accuracy in any particular setting, an approach that is commensurate with both the sought clarity of our agent-based framework and the absence of detailed knowledge regarding the properties of the collagenic tunic of \bschloss{}.

As well as being exposed to the elements, the vasculature of \bschloss{} exhibits a remarkable property, in that it periodically retracts and regrows its vessel network \cite{Madhu2020}. Whilst this behaviour can be externally stimulated \citep{Rodriguez2019}, it also occurs as part of the life cycle of the organism and serves as an example of its remarkable regenerative capability, with \bschloss{} capable of replacing excised vasculature within days \cite{Gasparini2014}. 
In addition, the ease with which the exposed vasculature and thin tunic can be imaged make \bschloss{} a valuable model organism for the investigation of tissue regeneration.
This repeated outgrowth of vessels through a persistent tunic also presents a platform for a broader question: to what extent can an extracellular medium, unmodified during vessel retraction, influence the formation of successive generations of vasculature? Such a property is not one that has been extensively explored in the context of chemotaxis-driven systems, though is reminiscent of hypotheses of wound healing and the accompanying angiogenesis \cite{Nardini2021,Flegg2020a}. With the tunic of \bschloss{} potentially facilitating such a persisting memory of the vessel network, we consider this question of long-term influence and remodelling in the context of our agent-based framework, revealing principles underlying dynamics of guided and temporally convergent vascular growth in response to mechanical cues.

Hence, in this study, we will develop a computational agent-based modelling framework that is tailored to exploring mechanically dominated vasculature development. Documenting and motivating our design choices in detail, with simplicity and interpretability in mind, we seek to capture phenomenology through a rich yet minimal ansatz for mechanical effects. Through a range of explorations, which evidence the ability of our model to reproduce familiar network structures, we highlight the potential for mechanical effects to give rise to diverse behaviours that are not typical of chemotactic systems. In particular, we observe that a persisting extracellular medium is capable of acting as a shared effective memory for periodically regrowing model vasculature, eliciting guided and evolving taxis without the presence of chemical signalling. Further, we see the emergence of a nuanced collective behaviour, with the clustered development of vasculature arising through a combination of redirection and remodelling, facilitated by mechanically mediated non-local interactions between developing vessels.

\section{A hybrid on/off-lattice model}
To model the growth and development of a small-scale vascular network, as found in \bschloss{}, we will employ an agent-based model, representing the apical tips of the growing vasculature as individual agents that evolve due to a specified set of rules, which may be coupled to their environment. These agents will lay a trail of tissue as they move, representing the constructed vessels and illustrated in \cref{fig: overview motion collision}A, a common approach in discrete models of vasculature development \citep{Stokes1991,Anderson1998,Scianna2013}. The behaviour of any given agent-based model depends significantly on the rule-set employed and the details of the implementation; below we motivate and describe the design choices made and the consequences that these decisions have on the emergent dynamics of the overall model. In doing so, we seek to overcome the often overwhelming complexity of agent-based models through clear and minimal design, providing a framework for hypothesis exploration that minimises the potential for 
artefactual dynamics specific to the modelling framework.
To achieve this, we will largely abstract away from the details of any given physical or biological system, seeking to capture general behaviours rather than produce quantitatively faithful measures of vascular development.

Motivated in part by the morphology of \botryschloss{}, we will assume that our agent-based model evolves within a planar annular domain, though this geometry can be simply generalised to accommodate vasculature development in three dimensions. Throughout, time will be treated as discrete, with the equations of motion for individual agents interpreted as finite difference approximations to continuous motility.
We will leave the link between discrete and continuous-time models in this framework for future studies.



\subsection{Motion and collision}
Motion in agent-based models is commonly split into one of two cases: on-lattice movement, where agents are confined to move on a preset grid, and off-lattice movement, where agents move unrestricted. Both approaches are associated with advantages and drawbacks, with off-lattice models requiring a potentially sophisticated scheme for the detection of collision between agents, whilst on-lattice models can easily suffer from rasterisation artefacts of the employed grid. Since both movement and collision are significant features of vascular development, and some implementation of collision is required for the formation of loops in vascular networks when cast as agent-based models, we seek to overcome the limitations of each of these schemes whilst benefiting from their distinct advantages.

Taking advantage of both on- and off-lattice approaches, we will employ a hybrid scheme, with an off-lattice component handling agent motion whilst an on-lattice scheme enables efficient calculation of collision events that occur between agents and any aspect of the constructed vasculature. In brief, agents (growing tips of the vasculature) move freely in the domain and are not restricted to movement on a grid. These tip positions are then separately rounded to a `collision grid' to facilitate straightforward determination of collision events.

In detail, with agents labelled via an indexing set $I = \{1,2,\ldots,N(t)\}$ at time $t$, the position of agent $i\in I$ is denoted by coordinates $(x_i,y_i)$ and is updated simply via the free-movement evolution equation
\begin{equation}\label{eq: position update rule}
    \begin{bmatrix}
    x_i(t+\dt) \\ y_i(t+\dt)
    \end{bmatrix} = 
    \begin{bmatrix}
    x_i(t) \\ y_i(t)
    \end{bmatrix} + 
    \begin{bmatrix}
    \cos{\theta_i(t)} \\ \sin{\theta_i(t)}
    \end{bmatrix}V\dt\,,
\end{equation}
where $\theta_i$ is the orientation of the agent relative to the fixed $x$-coordinate axis, $t$ is the current time, $V$ is the agent speed, assumed to be constant, and $\dt$ is a fixed timestep. In turn, the orientation is updated from one time to the next following
\begin{equation}\label{eq: orientation update rule}
    \theta_i(t+\dt) = \theta_i(t) + km + \sigma^2\xi\,.
\end{equation}
The term $m$ encodes the influence of mechanical effects on the direction of the agent, which we will later describe in detail in \cref{sec: mechanics}.
This factor is modulated by a weight $k\in[0,1]$, which limits the magnitude of the contribution of mechanical effects to the updated agent orientation. The final term, $\sigma^2\xi$, represents the contribution of a random noise term with zero mean and variance $\sigma^2$, which we take to follow a normal distribution, requiring that $\sigma^2$ scales with $\dt$. 
Though the form of this update rule is not significant, we note that it affords a notion of directional persistence to the agents, with $\theta_i$ unchanging in the absence of rotational noise or mechanical effects. Whilst this appears to be an intuitive supposition for the motion of the tips of vascular networks, relaxing this assumption is simple to accommodate in the modelling framework via the removal of the $\theta_i(t)$ term and a suitable adjustment to the additive directional noise. We defer this model modification to future investigations.

\begin{figure}[t]
    \centering
    \begin{overpic}[width=0.9\textwidth]{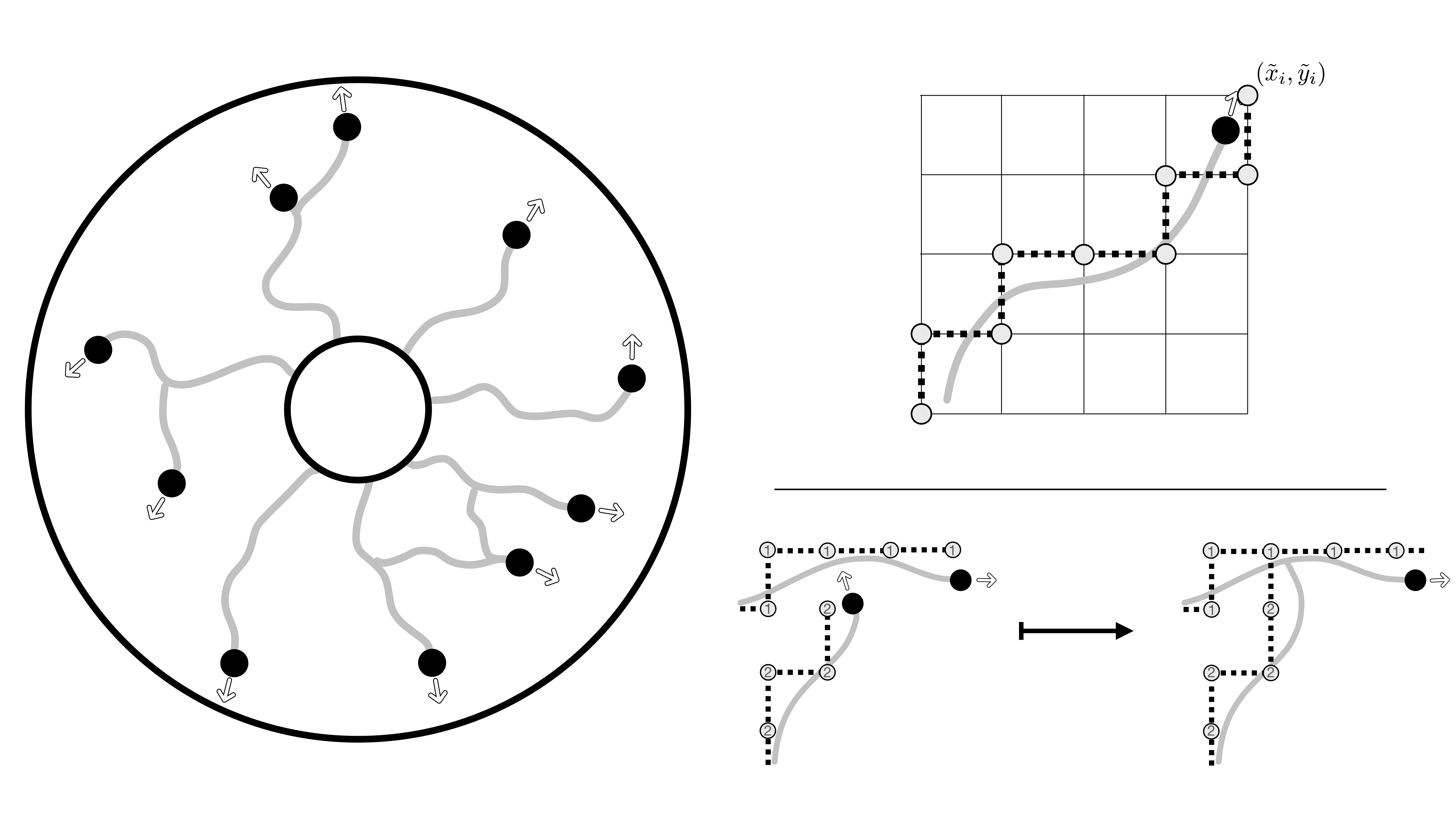}
    \put(0,52){(A)}
    \put(59,52){(B)}
    \put(49,20){(C)}
    \end{overpic}
    \caption{The agent-based model, the collision grid, and collision detection. (A) Illustrative evolution of vasculature from the inner edge of an annular domain, with the agents leaving behind a vessel trail as they move. Their motion in the domain is unrestricted, with instantaneous headings shown as outlined arrows. Agents are shown as black discs whilst trails are depicted as grey curves. (B) The rounding of off-lattice motion to the collision grid, with a smooth trajectory being rounded to a discrete path on the grid, the latter as illustrated by grey discs connected by dotted lines. Of note, this discrete path 
    is used only for the detection of collisions, stored alongside the unrestricted 
    path. (C) The detection of a collision via the collision grid, with the occupancy of vertices being queried when an agent attempts to move. In this example, the lower agent attempts to move close to a vertex previously occupied by the upper agent. This collision is resolved by connecting the separate vessel networks and removing the colliding agent from further simulation. Vertices of the collision grid are numbered by agents as they move, so that the grid records both the current position of the agents and the vessel trails laid during previous growth.}
    \label{fig: overview motion collision}
\end{figure}

With agent motion inherently unconstrained with this formulation, we define a mechanism to identify collisions between agents and obstacles, which can include both other components of the vasculature and any boundaries of the computational domain. Inspired by the natural convenience of on-lattice models for collision detection, we construct a relatively coarse Cartesian grid, onto which we round the agent positions, and denote these rounded values by $\tilde{x}_i$ and $\tilde{y}_i$. Of particular note, these rounded positions do not replace the unrounded coordinates used to track the position of the agents in \cref{eq: position update rule}, so that we do not inherit the limitations of purely on-lattice movement. As discussed further below, the spatial scale and geometry of the coarse grid can be chosen in a number of ways, potentially corresponding to natural scales in the biological system under investigation. With the rounded coordinates $(\tilde{x}_i,\tilde{y}_i)$ now belonging to a discrete set of positions, as illustrated in \cref{fig: overview motion collision}B, detecting a collision is as simple as checking if the grid location $(\tilde{x}_i,\tilde{y}_i)$ is occupied, either by another agent, the vessel trail left previously by an agent, or a boundary.

The discrete nature of this now-simplified collision problem lends itself to simple implementation, with only the previous rounded coordinates needing to be stored in order to enable rapid collision checking. The straight-line path travelled by the agent in the timestep is first queried on the collision grid, with any collisions identified and the appropriate tiles' occupancy updated. 
To enable agent-agent and agent-trail collision detection in a manner that identifies the agents or vessel trails involved, in practice we store the grid as an integer array, with agents marking their rounded paths via their index $i\in I$. An example collision event between an agent and a vessel trail is illustrated in \cref{fig: overview motion collision}C, highlighting the simplicity of our chosen scheme for collision detection. Of note, at each time step we randomise the order in which agents move and collisions are calculated in order to prevent agent-specific bias in the simulated network formation.

By combining free movement and rounding to a discrete collision grid, we circumvent the limitations of both on-lattice and off-lattice schemes, involving only the cost of rounding operations, storage of the lightweight collision grid, and rasterised collision detection. Notably, there remains substantial freedom in the specifics of the collision grid, with regular Cartesian grids being the easiest to implement whilst more complex grids, such as hexagonal grids, afford further freedom in grid shape. In the present study, we adopt a simple Cartesian grid, acknowledging that the details of the collision grid influence only the details of collision detection, rather than agent motion more generally. Further, there is no unique choice for the resolution of the collision grid, with refined grids giving rise to more-precise collision detection than coarse alternatives, though at increased computational costs. However, when the geometry of an agent can be approximated by a single grid tile, a natural interpretation is to conflate the agent size with the tile size. We will adopt this viewpoint in the remainder of this study, though the generalisation of this to complex agent morphologies is straightforward, with a refined grid capable of capturing the shape of an agent and, thus, collisions to any desired precision. In the broad context of existing agent-based models, this latter extension can be viewed as being in the spirit of the multi-site cellular Potts model \citep{Graner1992}, whilst the former resembles the approach of classical cellular automata.

\subsection{Splitting, merging, and inherent heterogeneity}\label{sec: splitting merging hetero}
A characteristic feature of vascular networks in many contexts is their complex branching structures, with individual vessels splitting off and subsequently merging with others to create intricate heterogeneous vasculatures. In order to replicate this common topology, we include minimalistic mechanisms for the splitting of agents and their merging with both each other and any vessel trails, once again seeking simplicity and phenomenology in our model and implementation to facilitate ready and interpretable exploration.

To capture regular but asynchronous branching in the model vascular network, we implement an agent-specific timer that determines when an agent will split in two, crudely mimicking the duration of a mitotic cycle within the cell, which has been modelled elsewhere and in other contexts with varying degrees of complexity \citep{Figueredo2013,Poleszczuk2016,VanLiedekerke2015,Ferrell2011}. At each timestep, the timer is decremented towards zero. Upon reaching zero, a new agent is created on a randomly selected neighbouring tile of the collision grid to the parent, recalling that the lengthscale of the collision grid may be interpreted as a proxy for the size of the agent, with appropriate generalisation for multi-tile agents. The initial value of each agent's timer is chosen uniformly at random and ranges up to a global maximum, $\thetaint$ so that splitting events are asynchronous but occur at prescribed intervals, though we note that the introduction of agent-specific mitotic periods presents a simple avenue for the incorporation of additional heterogeneity. After the creation event, there are many possibilities for the subsequent behaviours of both the parent and child agent, with particular biological contexts naturally informing the specification of such behaviours. Here, we proceed simply and abstractly, with both the parent and the child agent deviating from the initial heading of the parent by a preset splitting angle $\thetasplit$ in opposite directions. 
This choice is sufficient for generating qualitatively plausible network structures and provides a degree of freedom in specifying the angle of cell splitting. Recognising that many alternatives are possible, different behaviours may be readily implemented in place of this simple choice in the provided implementation, such as having the parent agent proceed unperturbed by the process of division. This also includes the clear potential for treating $\thetasplit$ as an agent-specific quantity or a random variable, which may contribute a further source of heterogeneity in future work.

Representing the opposite phenomenon to splitting and termed anastomoses in the context of vasculature, the merging or collision of agents, either with other agents or laid trails, is necessary to generate closed loops in planar vasculature, with tree-like networks otherwise being the only possibility. Noting that collisions with both agents and vessel trails can be efficiently and individually identified via the collision grid, it remains to describe how the behaviour of an agent is modified by a collision. Illustrated in \cref{fig: overview motion collision}C, the most simple and most frequent contact interaction occurs between an agent and a trail, which is resolved by merely deactivating the colliding agent and joining the two trails,
capturing the merging of a growing vessel with existing architecture. Other approaches, such as allowing the agent to pass through a trail, are not consistent with the assumed planarity of the network. A similar process is implemented when an agent collides with a boundary, with the agent simply becoming inactive. The comparatively rare event of agent-agent collision also proceeds analogously, with one of the agents being rendered inactive, though the subsequent behaviour of the remaining agent requires specification. 
Consistent without our aim of model flexibility, we allow for the easy specification of behavioural modification, such as the reorientation of the surviving agent to the mean heading of the colliding agents. In our illustrative examples of \cref{sec: exploring the model}, we opt to not alter the behaviour of the remaining agent at all, 
and our simulations that follow are insensitive to this choice.

There are many additional ways in which heterogeneity can be included in even this simplistic agent-based model of vasculature development, ranging from temporal evolution of agent characteristics to differences between individual agents, both of which are features of biological systems. In the present model, we have already remarked on the inclusion of qualitative effects of simple between-agent heterogeneity via the introduction of statistically independent directional noise and the distinct initial phases of the agent's timers that govern their splitting behaviours. In the next section, we will motivate and describe how each agent is coupled to its local environment, which is the most significant source of heterogeneity in the present modelling framework.

\begin{figure}
    \centering
    \begin{overpic}[width=0.9\textwidth]{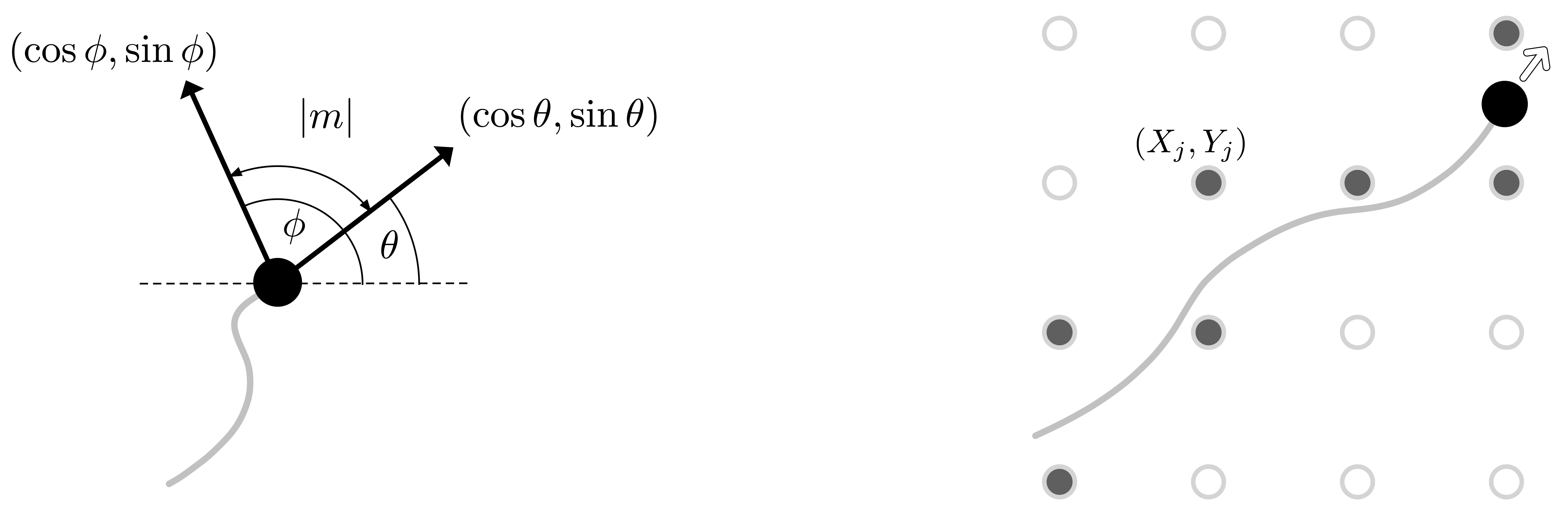}
    \put(-2,33){(A)}
    \put(58,33){(B)}
    \end{overpic}
    \caption{Updating the state of the tunic bed. (A) An illustration of the mechanical effect term $m(\theta,\phi)$, whose magnitude is the acute angle between the heading of the agent, $\theta$, and the current state of the bed, $\phi$. The heading and the preferred orientation of the bed are displayed as black arrows, annotated with the Cartesian components of the two directions. (B) The discrete grid $(X_j,Y_j)$ on which the basis functions are located is shown superimposed on the path of an agent as hollow circles, with those grid points closest to the trajectory, shown filled, having their weights $w_j$ updated following \cref{eq: weight update rule}.}
    \label{fig: intuition for mechanics}
\end{figure}

\section{Modelling mechanics}\label{sec: mechanics}
Drawing further inspiration from \botryschloss{} and its thin collagenic tunic bed, we model the effects of a surrounding material upon which the vasculature develops. This material will influence the direction of vessel growth, with the prominent direction of the collagen fibres acting as a guide for the agent motion. In the context of our implementation, this amounts to the local medium altering the direction of agent movement $\theta_i$. More precisely, we quantify the state of the medium by a single scalar quantity $\field$, which we interpret as the preferred orientation of fibres in the tunic bed, with $\phi$ being a function of both space and time. As illustrated in \cref{fig: intuition for mechanics}A and with reference to the relation of \cref{eq: orientation update rule}, we can now define the phenomenological mechanical effect term $m$ as a function of agent orientation $\theta$ and the field $\phi$ via
\begin{equation}
    \abs{m(\theta, \phi)} = \arccos{\abs{\cos{(\theta-\phi)}}}\in[0,\pi/2]\,,
\end{equation}
with the sign of $m$ being such that $\theta + m \equiv \phi \mod{\pi}$, which we note is uniquely defined. Here, we are implicitly evaluating $\phi$ at the location of the agent with orientation $\theta$, so that the mechanical response depends on the local state of the tunic bed. Of note, this response is invariant under both $\theta\mapsto\theta+\pi$ and $\phi\mapsto\phi+\pi$, so that the magnitude of $m$ is precisely the unsigned acute angle between the orientation of the agent and the preferred orientation of the bed. This also assumes a lack of directionality or polarity of the mechanical features, as might be associated with biopolymers embedded in the external environment.
Hence, the mechanical effect term $m$ captures a notion of non-alignment, as depicted in \cref{fig: intuition for mechanics}A. For completeness, we now state \cref{eq: orientation update rule} more precisely as
\begin{equation}\label{eq: orientation update rule full}
    \theta_i(t+\dt) = \theta_i(t) + km(\theta_i(t),\phi(x_i(t),y_i(t),t)) + n\xi\,,
\end{equation}
where $i\in I$ indexes the agents. In particular, if $k=1$, then $\theta_i(t+\dt) \equiv \phi(x_i(t),y_i(t),t) \mod{\pi}$ in the absence of noise, so that the motion of the agent aligns perfectly with the bed.

With $\field$ thereby influencing the direction of the agents moving on the tunic bed, the details of the feedback that agents have on the bed, and indeed the nature of $\field$ itself, must be specified. 
Here, we opt for an abstract approach that, rather than being tied to particular materials or phenomena, will afford significant flexibility and simple interpretation. 
In particular, inspired by the Green's function approaches that are common to linear differential equations, including the laws governing linear elasticity, we will suppose that $\phi$ may be written as a sum of basis functions $\basisfun$ at any given instant in time.
This assumes that a principle of spatial superposition exists in our idealised, phenomenological setting
and results in sufficiently rich dynamics to capture a qualitatively plausible notion of mechanical feedback, in line with the objectives of this study. In symbols, we define
\begin{equation}\label{eq: phi}
    \phi(x,y,t) = \sum_{j\in J}\weight_j(t) \basisfun(x,y,X_j,Y_j) + \phi_0(x,y)\,,
\end{equation}
where $J$ is a finite indexing set that labels the locations $(X_j,Y_j)$ of the basis functions, $\weight_j$ is the corresponding weight assigned to the $j$th basis function, and $\phi_0$ is the initial state. The locations $(X_j,Y_j)$ may be specified in a potentially unstructured manner, though here we simply distribute the basis functions on a uniform Cartesian grid, coarser than the collision grid by a factor of ten in each direction. We further approximate $\phi$ as being piecewise constant on a nearest-neighbour discretisation of the space via the grid of basis functions, so that $\phi(x,y,t) \approx \phi(X_j,Y_j,t)$, where $(X_j,Y_j)$ is the closest gridpoint to $(x,y)$. This affords significant computational efficiency, with the basis functions able to be pre-evaluated on the grid and linear combinations of these values calculated as needed during simulations. This efficiency is feasible due to the time dependence of the preferred orientation $\phi$ being encoded only in the weights $\weight_j$, with the $\weight_j$ being modified by the growth of nearby vessels and thereby linking the tunic bed to the movement of the agents.

Before specifying how the weights evolve over time, we first note that there is considerable freedom in the choice of basis function $\basisfun$. In particular, noting the diversity of linear partial differential equations, such as those of Stokes flow, this seemingly restrictive ansatz enables the inclusion of a range of effects and qualitative properties of the material bed. Here, hoping to explore non-local mechanical effects, we will assume the simple form
\begin{equation}
    \basisfun(x,y,X,Y) = \frac{1}{\lambda\left[(x-X)^2 + (y-Y)^2\right]+1}\,,
\end{equation}
where $\lambda$ determines the lengthscale over which the effects of this basis function decay. Our minimal choice indeed captures the notion of non-locality, with changes to the $\weight_j$ propagating throughout the domain. Notably, this function is non-singular, so that $\basisfun(X,Y,X,Y)$ is well defined, and it satisfies the condition $f(X,Y,X,Y)=1$, which we will assume throughout for notational convenience only. More generally, $\basisfun$ may readily be substituted for complex or biologically motivated forms in order capture particular effects, such as material anisotropy or longer-range impacts. Our minimalistic singularity-inspired approach bypasses the need for complex numerical solvers, thus affording significant efficiency whilst enabling qualitative freedom, though at the cost of quantitative accuracy.

Returning to the weights $\weight_j$, we will pose an update rule that captures our interpretation of $\phi$ as a preferred direction, such that if an agent moves along the direction specified by $\phi$, so that locally $\theta \equiv \phi \mod{\pi}$, then the state of the bed is unchanged and the corresponding weight is accordingly unmodified. Consistent with this, we will seek a rule where larger differences in orientation give rise to larger modifications to $\weight_j$, as is intuitive. Noting the duality between this sought response and the posed mechanical effects of the bed on agent motion, the response of the bed can be succinctly encoded via the same mechanical effect term, $m$. Explicitly, we pose
\begin{equation}\label{eq: weight update rule}
    \weight_j(t+\dt) = \weight_j(t) - \kappa m(\theta_i(t),\phi_j(X_j,Y_j,t))
\end{equation}
where $j\in J$ is the index of the nearest basis function to agent $i\in I$ and $\kappa\in[0,1]$ modulates the remodelling of the bed. In this formulation, $\kappa=1$ corresponds to perfect remodelling of the medium by the vasculature, whilst $\kappa=0$ represents a medium that is unaffected by vessel growth. In symbols, this entails that, when $\kappa=1$ and all other weights remain unchanged, we have $\phi(X_j,Y_j,t+\dt)=\theta_i(t) \mod{\pi}$, precisely mirroring the case of $k=1$ in the context of agent reorientation. In our implementation, acknowledging that an agent may pass close to multiple points $(X_j,Y_j)$ during its motion in a single timestep, this update rule is applied to all $w_j$ that are nearest to the agent at some instant within a timestep, as illustrated in \cref{fig: intuition for mechanics}b. With this update rule, we remark that our chosen form of $\phi$ does not converge to a limiting value as the resolution of the grid increases, as we have opted to parameterise our updates of the $w_j$ by the interpretable parameter $\kappa$. In particular, our assertion that $\kappa=1$ corresponds to perfect pointwise remodelling prohibits this notion of convergence. However, should convergence be desired in a given application, one may opt instead for an update rule that scales the weights inversely with the density of the grid, which readily yields convergence in the high resolution limit, though is not compatible with our interpretation of $\kappa$. 
Consistent with our goal of intuitive implementation and computational efficiency, we consider a fixed discretisation where convergence in this limit is not required.

\section{Exploring mechanical feedback}\label{sec: exploring the model}
We implement our agent-based model as described, freely available at \citep{botryllusCode}. In what follows, we will explore our minimal model of vascular development and a coupled mechanical bed in the context of \botryschloss{}, adopting an annular domain in line with the morphology of this model organism. In order to mimic observed growth dynamics of the vascular network, in what follows we will initialise agents on the inner boundary of the annulus, with initial headings aligned radially from the annulus centre. We will simulate growth until all agents have terminally collided with the laid vessel trails, the boundary of the domain, or one another. Typically, we will begin with between eight and twenty agents, with reference to the typical vasculature of \botry{}, though the explorations below are robust to variations in this initial condition as well as to differing refinements of the collision grid and $\phi$ discretisation, which we specify here as $1000\times1000$ and $100\times100$ uniform Cartesian grids, respectively. Full details of the parameters and initial conditions employed in each exploration below accompany the example implementation \citep{botryllusCode} and are summarised in \cref{app: parameters}.

\subsection{Mechanics-free network growth}
\begin{figure}
    \centering
    \begin{overpic}[width=0.9\textwidth]{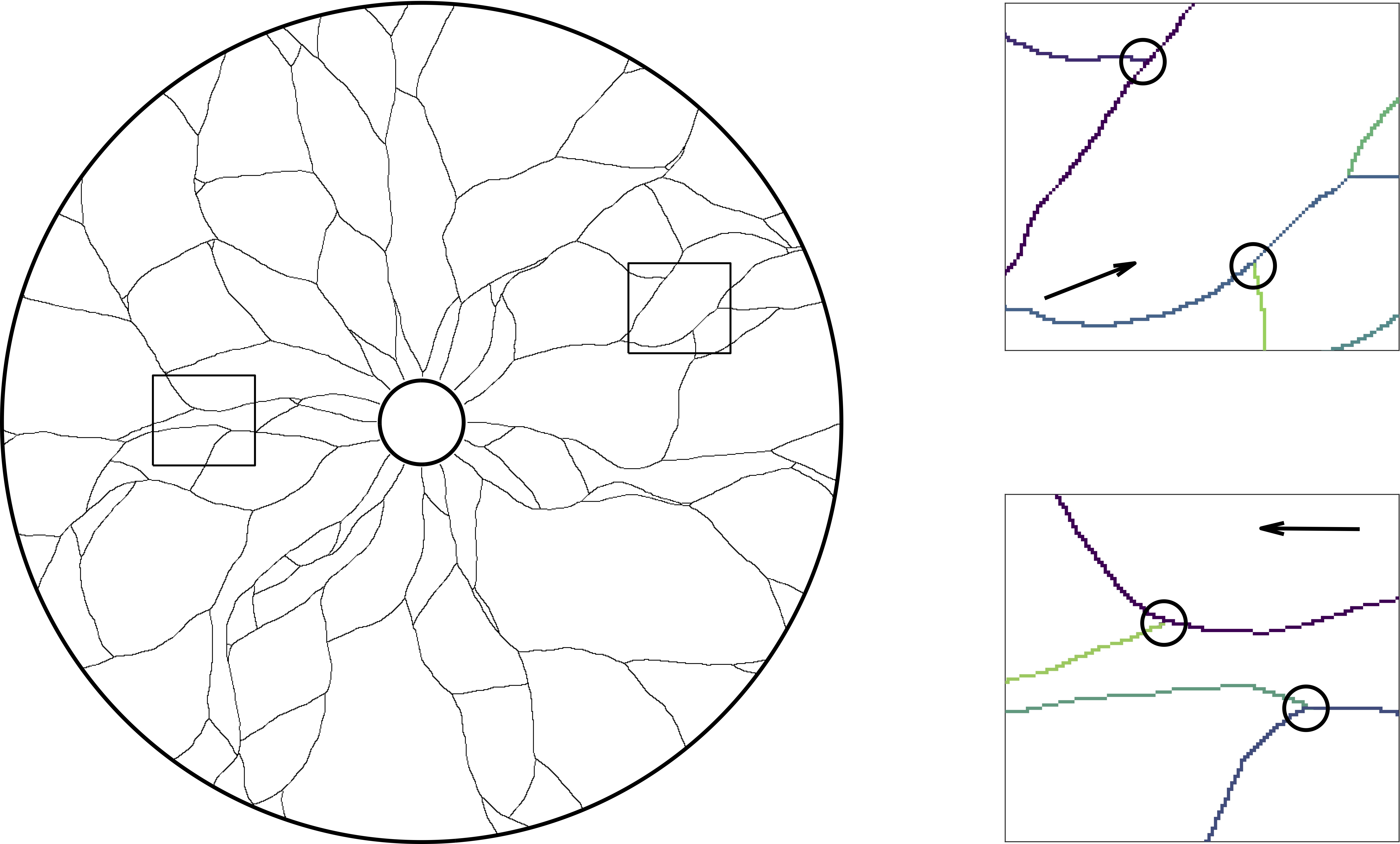}
    \put(0,61){(A)}
    \put(68,61){(B)}
    \put(68,26){(C)}
    \end{overpic}
    \caption{The growth of a network in an annulus in the absence of mechanical effects. (A) A simulated vascular network growing outwards from the inner edge of the annulus, in qualitative agreement with canonical network structure. Vessel trails are shown as thin black curves in this illustration, with all agents (not shown) having collided with vessel trails, the boundary of the domain, or one another. (B,C) The close-up vascular network,  with instances of (B) collisions and (C) splitting of agents, indicated by black circles and the paths of individual agents shown in distinct colours for clarity. The snapshots of (B) and (C) correspond to the outlined regions of the full network shown in (A), with the arrows displaying the general direction of agent movement during growth. Here, we have initialised agents facing radially outwards from the inner edge of the annulus and taken $k=\kappa=0$, with other parameters summarised in \cref{app: parameters}.}
    \label{fig: mechanics-free network}
\end{figure}

Firstly, as a qualitative verification of the basic behaviours of the framework, we simulate the development of a vessel network in the absence of any mechanical effects, equivalent to setting $k=\kappa=0$. The small scale of the \botry{}-inspired problem and the minimal nature of the model lends itself to rapid computation, with a single simulation typically taking $\SI{0.5}{\second}$ on standard personal desktop hardware (Intel Core i7-6920HQ CPU). In \cref{fig: mechanics-free network}A, we show a typical simulated vascular network via the collision grid, which we recall records the locations of the vessels and agents in the annular domain. The qualitative agreement between the emergent network structure and observed vasculature demonstrates the capability of minimal models to reflect observed networks, at least at a  phenomenological level. In \cref{fig: mechanics-free network}B and \cref{fig: mechanics-free network}C, we highlight occurrences of merging and splitting, respectively, colouring each vessel trail by the agent that laid it for clarity in these panels, which gives rise to the complex and interconnected structure seen in \cref{fig: mechanics-free network}A.

\subsection{Remodelling and periodic regrowth}\label{sec: remodelling and periodic regrowth}
\begin{figure}
    \centering
    \begin{overpic}[width=0.9\textwidth]{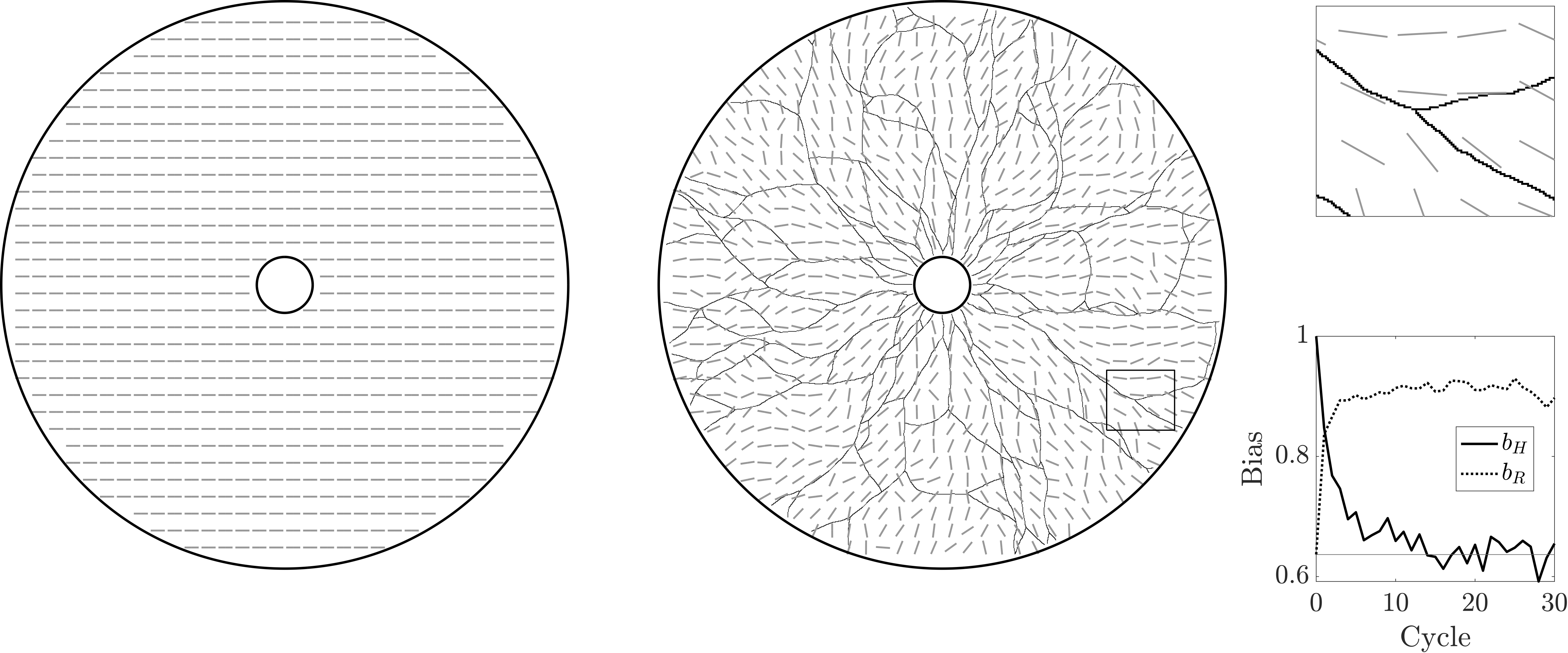}
    \put(0,42){(A)}
    \put(42,42){(B)}
    \put(80,42){(C)}
    \put(80,22){(D)}
    \put(15,3){Cycle 0}
    \put(56,3){Cycle 30}
    \end{overpic}
    \caption{The remodelling of the extracellular bed. (A) The initially uniform preferred orientation of the external bed is illustrated by grey oriented dashes, which are at an angle $\phi=0$ from the horizontal, sampled in the annular domain of vascular development. Of note, we do not distinguish between $\phi$ and $\phi+\pi$, a property captured by this manner of visualisation. (B) The state of the bed after 30 cycles of growth, overlaid with the grown vasculature. The initial horizontal configuration of the bed has been eroded by the repeated growth of the vasculature. (C) A close-up view of the rectangular region outlined in (B), highlighting the approximate local alignment of the bed with the vessel network. (D) Measures of overall bed alignment over multiple cycles of regrowth. The measure of horizontal bias, $b_H$, can be seen to quickly decay to the value expected for a uniformly distributed $\phi$, shown as a horizontal line, whilst the measure of alignment to a radial distribution, $b_R$, draws closer to unity, indicative of an approximately radial distribution of $\phi$. Here, we have taken $k=0$ and $\kappa=0.1$, with other parameters summarised in \cref{app: parameters}.}
    \label{fig: remodelling and periodic regrowth}
\end{figure}
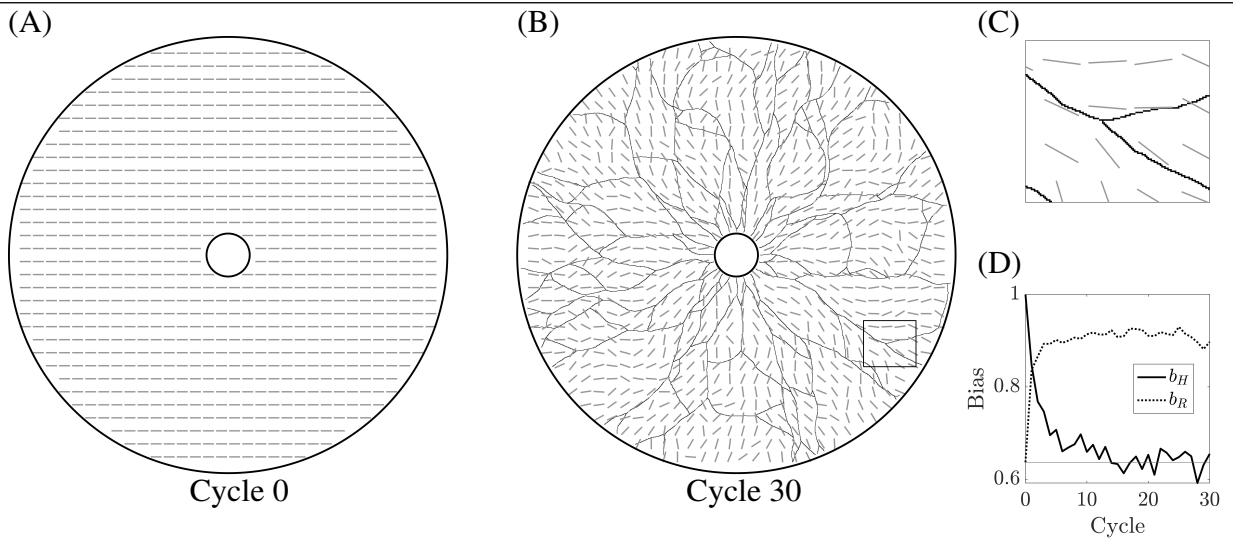

Having established 
qualitative validity of the most basic behaviours of the model, we now test the ability of the framework to capture a notion of agent-driven remodelling of the extracellular bed. To do so, we set $k = 0$ and take $\kappa\neq0$, so that the agents move independently of the bed whilst modifying its configuration. We begin with the bed in an initially aligned state, taking $\phi_0(x,y)=0$ everywhere to yield a \emph{horizontally} directed bed, as illustrated in \cref{fig: remodelling and periodic regrowth}A. Drawing inspiration from the periodic regrowth of vasculature seen in \botry{}, we then repeatedly simulate the growth dynamics of vasculature, beginning each new cycle with agents on the inner edge of the annulus. With $\kappa\neq0$, the agents modify the bed as they move, with these modifications persisting between each cycle and the external bed thereby acting as a form of persistent mechanical memory.

\Cref{fig: remodelling and periodic regrowth}B and \cref{fig: remodelling and periodic regrowth}C display the state of the bed after 30 cycles of growth, by which time the initial horizontal configuration has been eroded to a more disordered state. We can quantify this divergence from uniformity via a simple measure of bias $b_{H}$, defined for a given fully grown network as
\begin{equation}
    b_H \coloneqq \avg{\abs{\cos{[\phi(x,y) - 0]}}}\,,
\end{equation}
where $\avg{\cdot}$ denotes a spatial average. Computing this quantity discretely by sampling $\phi$ at $(X_j,Y_j)$ for $j\in J$, which we recall indexes the locations of the basis functions, we plot this measure of bias in \cref{fig: remodelling and periodic regrowth}D. With an initial value of unity, representing perfect horizontal bias, we see that this measure collapses approximately to $2/\pi$, the value expected of uniformly distributed $\phi$. Also shown in \cref{fig: remodelling and periodic regrowth}D is an analogous measure of radial bias, defined as
\begin{equation}
    b_R \coloneqq \avg{\abs{\cos{[\phi(x,y) - \arctan{(y,x)}]}}}\,,
\end{equation}
where, here, Cartesian coordinates $(x,y)$ are such that the centre of the annulus is at $(x,y)=(0,0)$ and $\arctan$ is the four-quadrant inverse tangent. Of note, $b_H=1$ implies that $\phi\equiv0 \mod{\pi}$, whilst $b_R=1$ occurs only when $\phi$ is purely radial. Hence, the approach of $b_R$ towards unity and $b_H$ away from unity over many cycles of growth suggests a notion of convergence of $\phi$ away from the horizontal state towards a radial state, as might be expected given the radial initial headings of the agents and the directional persistence included in our formulation.

\subsection{Mechanics-dominated growth}\label{sec: mechanics-dominated growth}
\begin{figure}
    \centering
    \begin{overpic}[width=0.9\textwidth]{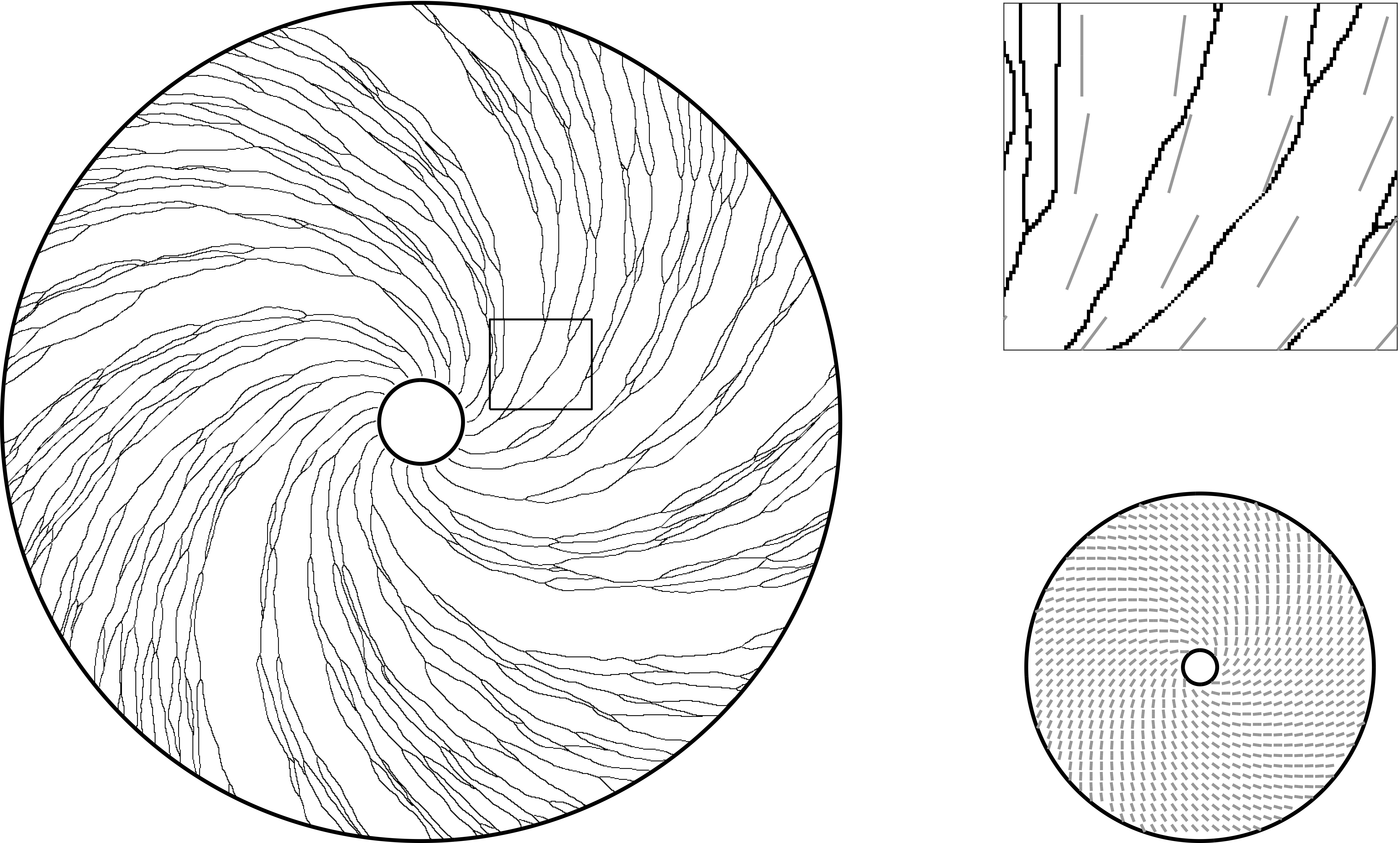}
    \put(0,61){(A)}
    \put(68,61){(B)}
    \put(68,26){(C)}
    \end{overpic}
    \caption{The mechanics-dominated growth of a network. (A) The developed vasculature, whose growth was strongly influenced by a spiral-like extracellular bed of fixed configuration. Accordingly, the vessel network resembles an anticlockwise spiral, though we note that noise and splitting events give rise to a complex network structure. (B) A close-up view of the rectangular region outlined in (A), with the fixed state of the extracellular bed shown as grey dashes, which represent the preferred direction of the bed. (C) The initial and unchanging state of the model tunic bed, which spirals anticlockwise and outwards from the inner edge of the annular domain. Here, we have taken $k=0.3$ and $\kappa=0$, with other parameters summarised in \cref{app: parameters}.}
    \label{fig: mechanics dominated growth}
\end{figure}

We now consider the opposite case to the above, one in which the tunic bed influences the development of the vasculature but is itself static, corresponding to $\kappa=0$ and $k\neq0$. In \cref{fig: mechanics dominated growth}, we illustrate the effects of an outwardly spiralling field $\phi$, which significantly alters the growth of the vessels, having taken $k=0.3$ in this example. The impacts of the spiral-like structure of the extracelluar bed, itself depicted in \cref{fig: mechanics dominated growth}C and given by $\phi(x,y,t) = \phi_0(x,y) = \arctan( -x-y, y-x)$, are readily observable in \cref{fig: mechanics dominated growth}A, with the vessel network displaying a vastly similar character to the influencing external medium. Despite the strong coupling between the bed and the growing vessels, the distorting effects of noise and agent splitting still serve to generate an intricate structure. Nevertheless, the grown vessels are approximately aligned with the preferred heading encoded in $\phi$, as can be seen in the close-up view of \cref{fig: mechanics dominated growth}B, where the direction of the extracellular medium is shown as grey dashes. Further increasing $k$ results in networks that even more closely track the spiralling field, though the formation of loops via collisions is inhibited by the strongly 
prescribed motion.

\subsection{Regrowth, remodelling, and mechanical influence}\label{sec: regrowth remodelling and mechanical influence}
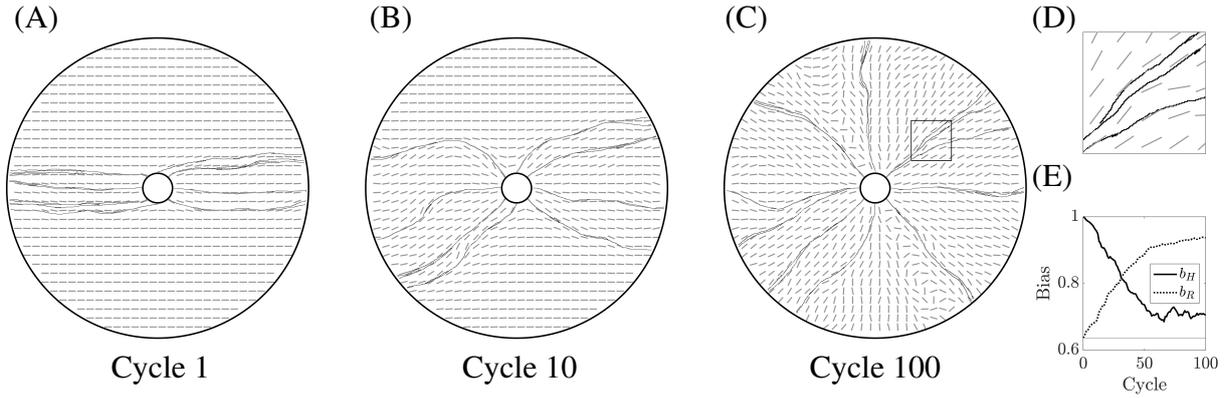
\begin{figure}
    \centering
    \begin{overpic}[permil,width=0.9\textwidth]{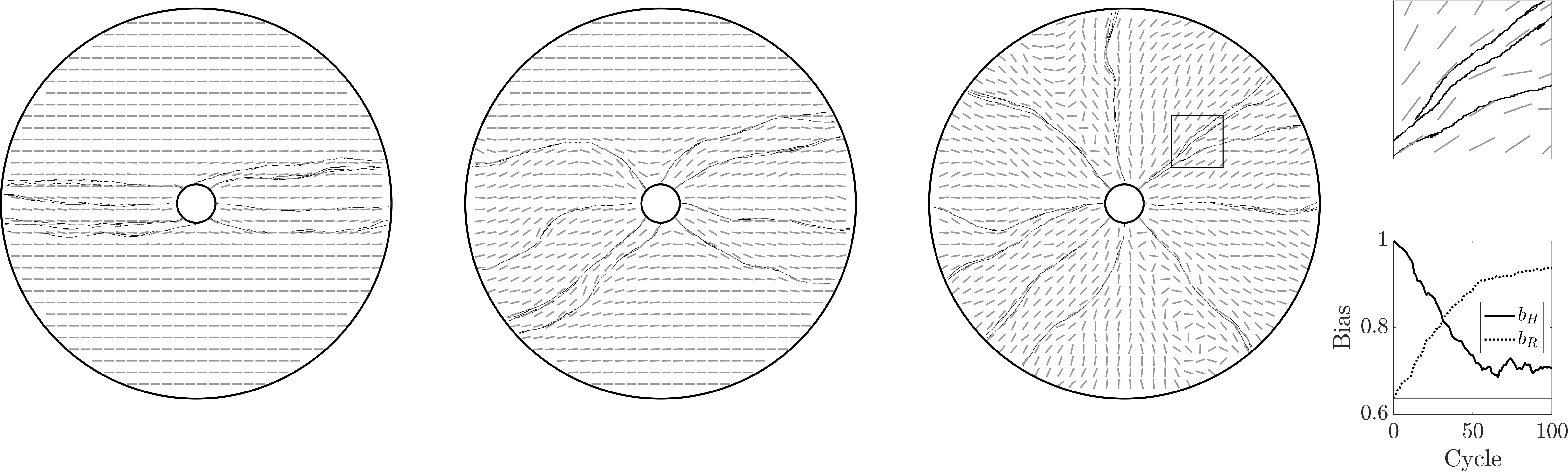}
    \put(10,457){(A)}
    \put(450,457){(B)}
    \put(890,457){(C)}
    \put(1270,457){(D)}
    \put(1270,260){(E)}
    \put(130,20){Cycle 1}
    \put(565,20){Cycle 10}
    \put(995,20){Cycle 100}
    \end{overpic}
    \caption{Mechanically coupled cyclic regrowth of a vessel network. (A,B,C) Beginning with a horizontally aligned bed, as in \cref{fig: remodelling and periodic regrowth}a, we simulate 100 cycles of vascular regrowth, with the vessels both modifying and being influenced by the tunic bed. We illustrate the grown networks and the state of the bed after 1, 10, and 100 cycles of growth in (A), (B), and (C), respectively. The initial configuration of the bed has a clear influence on the grown vasculature in (A), though the persistence of the initial state wanes over successive cycles of growth, with (C) approximately resembling a radially configured bed. (D) A close-up view of the rectangular region outlined in (C), from which we note that vessels are indeed splitting, though the strong coupling of growth direction to the tunic bed ensures that neighbouring vessels grow almost parallel to one another. (E) Measures of overall bed alignment over multiple cycles of growth, showing a sharp decrease in horizontal bias and a steady increase in radial bias, consistent with the erosion of the horizontally aligned initial state of the bed. The value of bias expected for a uniformly distributed $\phi$ is shown as a horizontal line. Here, we have taken $k=0.8$ and $\kappa=0.3$, with other parameters summarised in \cref{app: parameters}.}
    \label{fig: regrowth remodelling and mechanical influence}
\end{figure}

Having tested and illustrated the various behaviours and phenomenology captured by our model in relative isolation, we investigate 
a scenario where the tunic bed and the vasculature are fully coupled, taking $k\neq0$ and $\kappa\neq0$. In particular, adopting the horizontally aligned initial field $\phi$ of \cref{fig: remodelling and periodic regrowth}A, in \cref{fig: regrowth remodelling and mechanical influence} we explore the evolving vessel networks generated over successive cycles of growth, capturing how the extracellular medium is altered with each iteration. Here, we have taken $k=0.8$ and $\kappa=0.3$, which represent strong mechanical influence and moderate remodelling, respectively. In \cref{fig: regrowth remodelling and mechanical influence}A, we show the vasculature after a single cycle, from which the guiding effects of the bed are evident in the broadly horizontally aligned network. Of note, after this single cycle of growth, the state of the medium is no longer perfectly aligned to the horizontal, with the vessels having altered the nearby weights $w_j$ during their growth. By the tenth cycle, shown in \cref{fig: regrowth remodelling and mechanical influence}B, the bias of the horizontal initial state of the tunic is still visible, but has been significantly eroded by the repeated stochastic growth of the vessel network. At the conclusion of the final cycle of growth, illustrated in \cref{fig: regrowth remodelling and mechanical influence}C, the influence of the initial condition has all but vanished, with the grown network appearing approximately radial, similar in alignment to that of \cref{fig: remodelling and periodic regrowth}B. However, in contrast to \cref{fig: remodelling and periodic regrowth}B, the strong coupling of the vessel growth to the state of the tunic bed entails that neighbouring vessels grow in approximately parallel directions, leading to tightly packed vasculature, as highlighted in the close-up view of \cref{fig: regrowth remodelling and mechanical influence}D. Quantitatively, the waning of the initial horizontal alignment of the bed is captured by the bias $b_H$, illustrated in \cref{fig: regrowth remodelling and mechanical influence}E alongside $b_R$, which also demonstrates approximate convergence towards radial alignment.

\subsection{Non-locality and flocking}\label{sec: nonlocality and flocking}
In the above, we've seen how the model tunic bed can significantly influence the growing vasculature, as well as how the development of vessels can remodel the substrate. Here, in our final exploration, we will highlight a behaviour that emerges due to the inherent non-locality of the mechanical bed, which is difficult to discern qualitatively in the previous examples. Indeed, thus far, the major impact of the non-local form of $\phi$ in \cref{eq: phi} has been only to render the directional field of the tunic as a smooth function of space, which features in all of the previous examples.

\begin{figure}
    \centering
    \begin{overpic}[permil,width=0.9\textwidth]{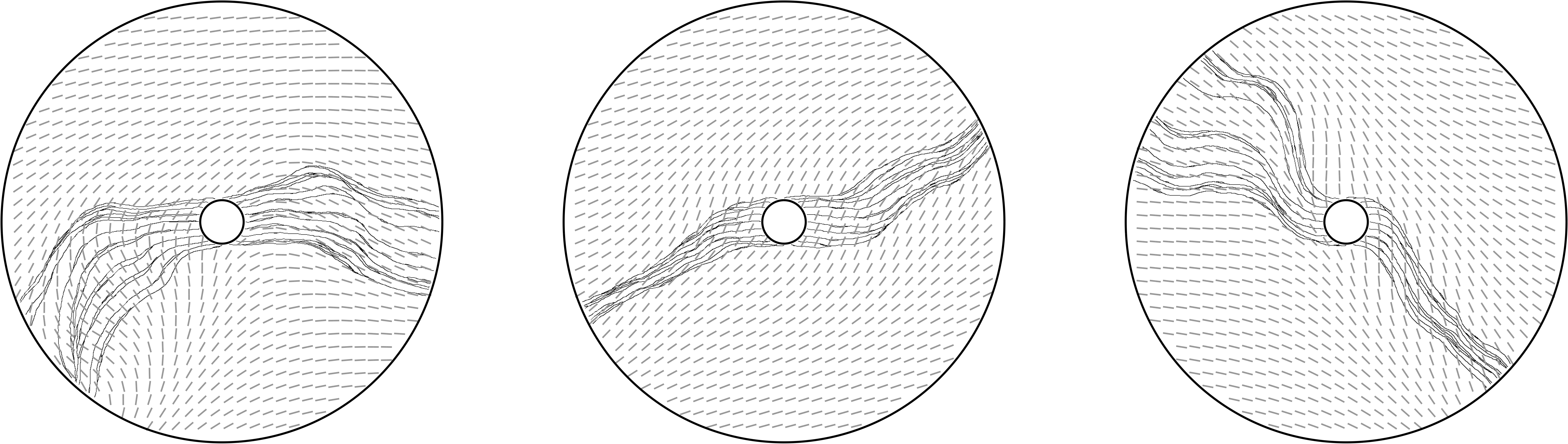}
    \put(0,427){(A)}
    \put(548,427){(B)}
    \put(1080,427){(C)}
    \end{overpic}
    \caption{Mechanics-mediated collective behaviour. (A,B,C) Three vascular networks grown from the same initial state, with $\phi_0=0$, dominated by the long effective range of modifications to the mechanical bed. With non-locality enhanced compared to previous explorations, an emergent collective behaviour is visible, with vessels closely following the paths of their neighbours due to the coupling to the tunic bed. The potential for nuance, as opposed to a uniform global behaviour, is evidenced in (A), where `flocks' of agents move in non-colinear directions, whilst a complex splitting of flocks is visible in (C).  Here, we have taken $k=0.8$, $\kappa=0.8$, and $\lambda = 10$, the latter greatly decreased from previous explorations, with other parameters summarised in \cref{app: parameters}.}
    \label{fig: nonlocality and flocking}
\end{figure}

In particular, we will consider the same mechanical setup as that of \cref{fig: remodelling and periodic regrowth}, with $\phi$ initially corresponding to a perfectly horizontal alignment. However, with reference to the parameters listed in \cref{app: parameters}, we now increase the effective range of the basis functions $f$, which contribute to $\phi$ via \cref{eq: phi}, by an order of magnitude, seeking to emphasise the potentially significant role of non-locality. In \cref{fig: nonlocality and flocking}, we illustrate multiple instances of growth, beginning with 20 agents on the inner boundary of the annular domain. Each of these examples serves to highlight an emergent collective behaviour, which we term flocking, where the vessels have grown in markedly similar but evolving directions, crudely reminiscent of the canonical collective motion of birds or fish. Here, this behaviour has emerged as the result of agents modifying the local state of the tunic bed, changes which then propagate to neighbouring vessels that, in turn, align along the new direction. Further, we see that the effects on behaviour are nuanced and are not global, with the vasculature on the left and right sides of \cref{fig: nonlocality and flocking}A having evolved in non-colinear directions. Additionally, we see that local branching can cause the separation of `flocked' vessels, as can be seen in \cref{fig: nonlocality and flocking}C. Hence, we observe that the inherent potential for non-locality in our framework is capable of giving rise to complex and emergent collective behaviours, mediated by the non-local mechanics of the bed.

\section{Discussion}
In this study, we have described, implemented, exemplified, and explored an agent-based model for simulating the development of mechanically dominated vascular networks. Noting the common complexity and intricacy of agent-based modelling frameworks in mathematical biology, we have sought to utilise minimal and simple schemes to realise phenomenology in each aspect of our model, which we have detailed, motivated, and illustrated throughout. In addition to facilitating independent reproduction and verification of model outputs, this simplicity has afforded significant computational efficiency and interpretability to our framework. This latter feature is evident throughout \cref{sec: exploring the model}, with the behaviours observed in our explorations being easily and readily relatable to the specific parameter choices and set-up of each example, with confounding factors minimised. As such, we expect that a primary future utility of the model will be in exploration and qualitative hypothesis testing, with the links between cause and effect not obfuscated by complex underlying systems or design choices.

Our ethos of seeking simplicity and phenomenology also extended to the primary focus of this work, which has been the introduction of mechanical guidance cues in place of chemotactic signals, the latter being appropriately commonplace in vascular models. Motivated in part by the model organism \botryschloss{}, whose morphology and environment suggests a subdominance or absence of chemical effects, we have sought to include the effects of a guiding external substrate on the growth of the network, with the directions pursued by the vessels being influenced by the state of a mechanical tunic bed. In posing our model of the mechanical bed, we have pursued phenomenology, rather than be tied to the properties of a particular material, resulting in a simple superposition-based model that nevertheless captures rich effects that are qualitatively plausible for a mechanical substrate. This model also lends itself to a computationally efficient implementation, in principle further enabling the rapid and lightweight enquiry that is well suited to exploratory study, at least at the level of qualitative effects.

In exploring the consequences of our modelling choices, we have encountered features and behaviours that might not be expected of networks shaped by chemotaxis. For instance, in \cref{sec: regrowth remodelling and mechanical influence}, the state of the tunic bed served as an effective memory for the periodically regrowing vasculature, which is highlighted in the extreme example of \cref{sec: mechanics-dominated growth}. These examples illustrate the ability of the proposed model to capture an intuitive notion of an environmentally shaped network, one in which mechanics guides the evolution of the network structure. Further, in our final examples, we have also seen how agents can erode and reshape the shared external medium, with the evolving network altering the environment and, in turn, the evolution of future vasculature on the same substrate. In the context of the periodically regrowing vasculature of \botry{}, this prompts future evaluation of the hypothesis that such a mechanical effect can give rise to an approximately convergent network structure, noting in particular the approximate convergence of the tunic bed in \cref{fig: remodelling and periodic regrowth}.

Perhaps most remarkable amongst our observations is the flocking behaviour seen in \cref{sec: nonlocality and flocking}, wherein nearby agents move in a collective fashion. Though such a clustered flocking might be expected of chemically driven development, with nearby agents sensing similar chemical cues, here this collective behaviour emerges via mechanics alone, with the non-local nature of the bed acting as a form of inter-agent communication. Hence, this suggests that chemical cues are not required for the realisation of coordinated growth dynamics, with mechanically inspired non-local directional cues able to evoke this comparable behaviour \textit{in silico}. 
We have also seen that there is considerable nuance to these mechanical effects, in that there still exists a balance between non-local cues and local, agent-level effects, as illustrated in the splitting and diverging flocks of \cref{fig: nonlocality and flocking}C and the non-colinear trajectories of \cref{fig: nonlocality and flocking}A, for instance.

In this initial work, our explorations have been limited both to two dimensions and to a particular, isotropic form of mechanical basis function. Each of these aspects represents a broad avenue for future generalisation and enquiry, with the introduction of material anisotropy through varied basis functions likely being a route for further development and diverse exploration. The consideration of network growth in three dimensions, as is more faithful to many biological circumstances, overall represents a minor modification to the proposed framework, though due care is needed in the treatment of agent orientation in three dimensions. Nevertheless, the two-dimensional examples explored in this work remain of pertinence to model organisms, whilst representing a natural testbed for the initial study of the impacts of mechanical effects on network development. Further, within this constraint, there is considerable scope for the exploration of more complex and alternative domains for vascular growth, noting that it is simple to generalise the annular \botry{}-inspired domain studied in this work. As an example, this could include the study of simulated excision experiments, where regions of the tunic bed of \botry{} are surgically removed and subsequently regenerated by the organism \cite{Gasparini2014}, which represents a promising direction for future enquiry into mechanically influenced vessel development.

In summary, we have presented and exemplified a simple computational model for the exploration of the role of mechanics in the growth of vascular networks. Throughout, we have opted to make minimal, intuitive design choices that facilitate ready interpretation and rapid simulation, which has enabled us to explore a variety of examples and regimes. In doing so, we have seen how a phenomenologically modelled mechanical substrate can act as an effective and malleable memory for growing vasculature, giving rise to complex and evolving network structures that are reminiscent of canonical vessel networks. Further, we have evidenced the potential for mechanical effects to perform a role typically associated with diffusive chemical species, here resulting in the emergence of complex, mechanically mediated collective behaviours.

\appendix

\section{Simulation parameters}\label{app: parameters}
The dimensionless parameters used in the generation of the figures shown in the main text are given in \cref{tab: parameters}, and can also be found accompanying the provided implementation \cite{botryllusCode}. In all simulations, we have adopted an annular domain of inner and outer radii of 0.1 and 1, respectively. The Cartesian grids for the collision grid and the basic functions are at a resolution of $1000\times1000$ and $100\times100$, respectively, each distributed in the square domain $[-1,1]\times[-1,1]$.

\begin{table}
\renewcommand{\arraystretch}{1.2}
    \centering
    \begin{tabular}{|r|r|r|r|r|r|l|}
    \hline
        Parameter &  \cref{fig: mechanics-free network} & \cref{fig: remodelling and periodic regrowth} & \cref{fig: mechanics dominated growth} & \cref{fig: regrowth remodelling and mechanical influence} & \cref{fig: nonlocality and flocking} & Description\\\hline
        $k$ &           0  & 0    & 0.3                 & 0.8  & 0.8 & Mechanical influence\\
        $\kappa$ &      0  & 0.1  & 0                   & 0.3  & 0.8 & Remodelling strength\\
        $\lambda$&      -  & 1000 & 1000                & 1000 & 10  & Locality of mechanics\\
        $N(0)$ &        16 & 20   & 20                  & 8    & 20  & Initial number of agents\\\hline
        $\dt$ & \multicolumn{5}{c|}{1}                               & Timestep\\
        $V$ & \multicolumn{5}{c|}{0.01}                              & Agent speed\\
        $\sigma^2$ & \multicolumn{5}{c|}{0.01}                       & Variance of noise\\
        $\thetaint$ & \multicolumn{5}{c|}{25}                        & Splitting period\\
        $\thetasplit$ & \multicolumn{5}{c|}{$\pi/6$}                 & Splitting angle \\\hline
    \end{tabular}
    \caption{Simulation parameters. We document the dimensionless simulation parameters used to generate the examples shown in the figures throughout the manuscript, which are also provided with the accompanying implementation \cite{botryllusCode}. Parameters in the lower section of the table are fixed in all simulations.}
    \label{tab: parameters}
\end{table}


\section*{Conflict of Interest Statement}

The authors declare that the research was conducted in the absence of any commercial or financial relationships that could be construed as a potential conflict of interest.

\section*{Author Contributions}
B.J.W. and A.T.D. contributed to the conception and design of the study, performed the research, and wrote and critically revised the manuscript.

\section*{Funding}
B.J.W. is supported by the Royal Commission for the Exhibition of 1851 and by the UK Engineering and Physical Sciences Research Council (EPSRC), Grant No. EP/R513295/1, EP/N509711/1.
A.T.D. is supported by the National Institute of General Medical Science of the National Institutes of Health (NIGMS-NIH) under award number 1R01GM132651-01A1. 

\section*{Acknowledgments}
The authors thank Megan Valentine and Younghoon Kwon for discussions regarding \botryschloss{} and for generously providing the image of \cref{fig: botryllus}.


\section*{Data Availability Statement}
The computational model generated in this study, along with all employed parameter sets, can be found at \cite{botryllusCode}.

\bibliographystyle{frontiersinHLTH&FPHY} 

\bibliography{references}



\end{document}